\documentclass[]{article}

\newcommand{\rs}{r_{\rm s}}
\newcommand{\bilder}{.}
\newcommand{\cuwi}{7.8cm}
\newcommand{\tuwi}{7.cm}
\newcommand{\cuwitwo}{12.cm}
\newcommand{\bhwi}{7.cm}

\newcommand{\NC}{Nuovo Cimento}
\newcommand{\PR}{Phys. Rev.}

\usepackage{epsfig}
\usepackage[labelsep=period]{caption}

\usepackage[latin1]{inputenc}
\usepackage[T1]{fontenc}
\usepackage{pslatex}

\begin{document}

\vspace*{0.8cm}
\noindent
{\LARGE\sffamily\bfseries First-person visualizations 
	   of the\\ special and general theory of relativity}

\bigskip
\bigskip
\hfil\begin{minipage}{13.0cm}
{\large\sffamily\bfseries U Kraus}

\bigskip
{\small Theoretische Astrophysik,
Universit\"at T\"ubingen,
Auf der Morgenstelle 10 C, 
D-72076 T\"ubingen, Germany

\smallskip
E-mail: ute.kraus@uni-tuebingen.de}

\bigskip
\smallskip
\today

\bigskip
Accepted for publication in The European Journal of Physics

\bigskip
\smallskip
{\normalsize\sffamily\bfseries Abstract}

Visualizations that adopt a first-person point of view
allow observation and, in the case of interactive simulations,
experimentation with relativistic scenes. This paper
gives examples of three types of first-person visualizations: 
watching objects that move at nearly the speed of light, 
being a high-speed observer looking at a static environment
and having a look-around near a compact object.
I illustrate and explain the main aspects of the visual observations,
outline their use in teaching relativity and report on
teaching experiences.
For teaching purposes, our visualization work is available on the web site
www.spacetimetravel.org and its German counterpart
www.tempolimit-lichtgeschwindigkeit.de.
This paper assumes some basic knowledge about relativity
on the part of the reader. It addresses instructors of physics
at the undergraduate and advanced secondary school level
as well as their students.

\medskip\noindent
(Some figures in this article are in colour only in the electronic version)
%(In the electronic version figures 1, 4, 7 and 8 are in colour)

\end{minipage}

%\maketitle

\vspace*{0.5cm}

\section{Introduction}

The theory of relativity started out with the reputation of being 
a particularly difficult and abstract theory. 
Anecdotes about this topic are widely known such as
Einstein's question `Why is it that nobody understands me, 
yet everybody likes me?'\footnote{Einstein, 1944.}
or the story about Sir Arthur Eddington
who, when somebody called him one of the three men 
in the world who really understood the theory of relativity,
replied that he did not know who might be the third.

Today, relativity theory is widely taught at universities
and increasingly finds its way into secondary school curricula.
Roman Sexl analysed this transition from a seemingly incomprehensive
theory to a widely taught one (Sexl 1980) and explained it by 
increased faith in the theory and better verbal explanations of
the relevant effects. 
His conclusion in 1980 was 
that for further progress in teaching relativity,
suitable audiovisual media should be developed.

In this contribution I will focus on one type of 
visual teaching aides, namely visualizations that adopt
a first-person point of view.
Many difficulties in understanding the theory of relativity
come from the fact that relativistic effects are not part of
everyday experience and may even seem to contradict this experience.
Also, direct observation by way of demonstration experiments 
generally is not possible. First-person visualizations are
meant to fill this gap. Simulated movies allow `observations'
and interactive computer simulations provide virtual experience
of relativistic scenes. 

The results of these simulations are often perplexing, even for
experts in the field. This is well illustrated by early
textbook illustrations of the visual appearance of objects
moving at nearly the speed of light: they were wrong
(Gamow 1940 (corrected in Gamow 1961), Fuchs 1965). 
Only from 1959 onwards, 
a series of investigations clarified what such an object would
really look like (Penrose 1959, Terrell 1959; an earlier study
by Lampa (1924) remained largely unnoticed). 
The detailed explanation of the phenomena by means of sketches
and animations is therefore essential for teaching purposes,
and we put equal emphasis on computing the simulations and 
on providing explanatory material.

In the following, I will illustrate the main aspects of
these first-person visualizations with several examples, 
outline their use in teaching relativity and report
on teaching experiences in university and secondary school.

\section{Visualizations of the special theory of relativity}

Special relativistic effects are minute in everyday life,
but become dramatic when velocities close to the speed of light
are involved.
The visualizations therefore explore high-speed motion. This
is done from two different points of view: we either watch
high-speed objects moving by or we move ourselves 
at nearly the speed of light through a static scene.

By the principle of relativity, these two points of view are
perfectly equivalent, i.e.\ given the same relative velocity,
the resulting movie is the same. 
The explanation, however, is quite different in the two different
reference frames, as outlined below: 
the flight times of light signals
explain the visual effects
for high-speed objects, while high-speed motion of the observer
provides a striking illustration of the aberration of light.

There are three different approaches to the choice of
length and time scales as follows. 
Either the objects have dimensions of the order of 
light seconds
like stars or large planets,
then the virtual fly by takes some seconds and can comfortably 
be recorded by a camera. 
Or the scene is taken from everyday life, e.g.\ houses or
trains, in which case the fly by time would be extremely short. 
We can then either invoke an improbably fast high-speed camera 
(in line with the equally improbable high-speed motion) 
and play the film in extremely slow motion
or, following Gamow (1940), we can simulate a virtual world
with a suitably reduced speed of light of the order of everyday
velocities.
In our examples we freely use either stellar or terrestrial objects,
as needed to make the respective simulation both appealing and instructive.

\subsection{Moving object}

%1
\begin{figure}
\begin{center}
%(a)\quad\epsfig{file=\bilder/w_0_2.ps,width=\cuwi}
(a)\quad\epsfig{file=\bilder/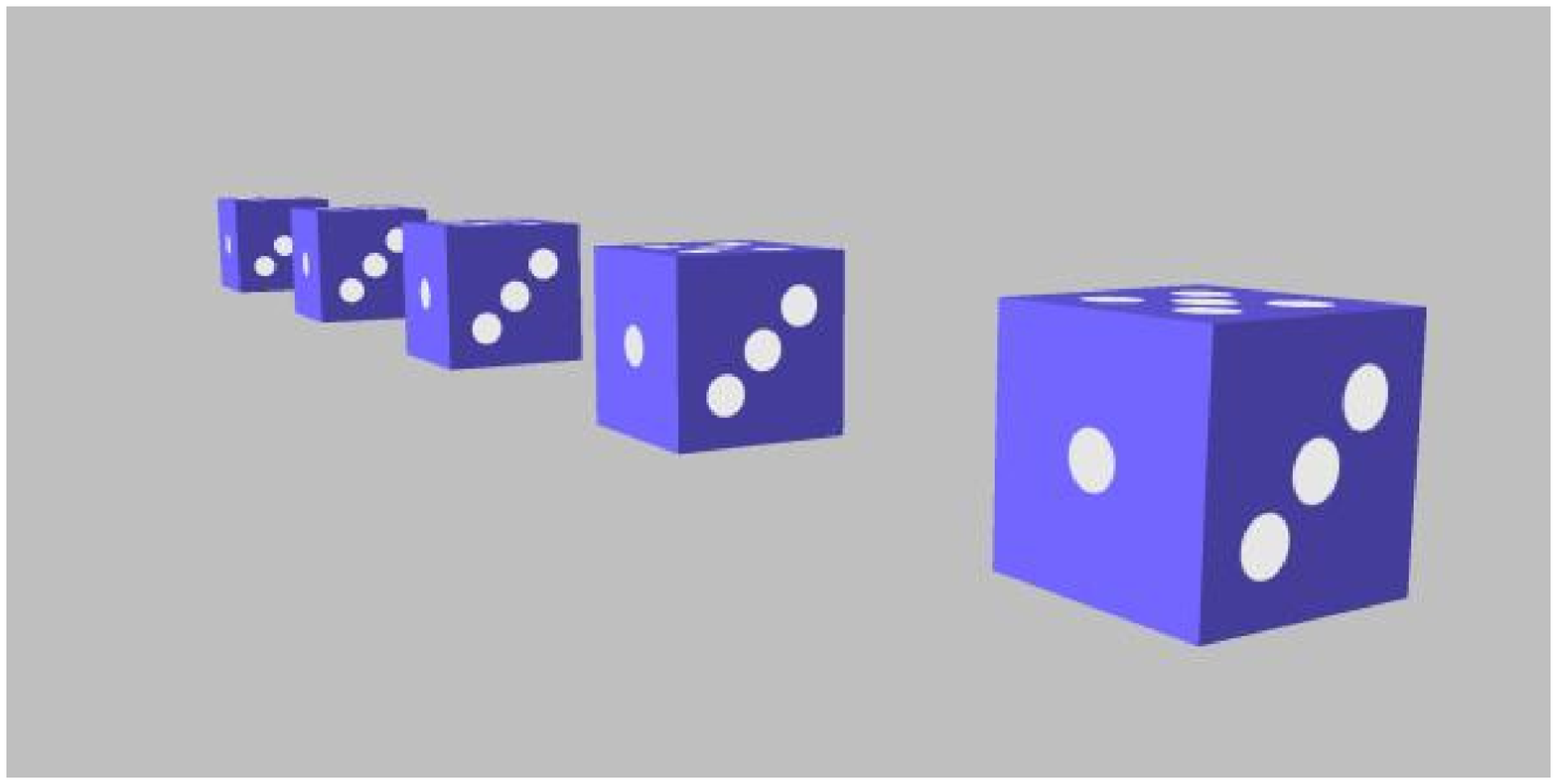,width=\cuwi}

%(b)\quad\epsfig{file=\bilder/w_3_2.ps,width=\cuwi}
(b)\quad\epsfig{file=\bilder/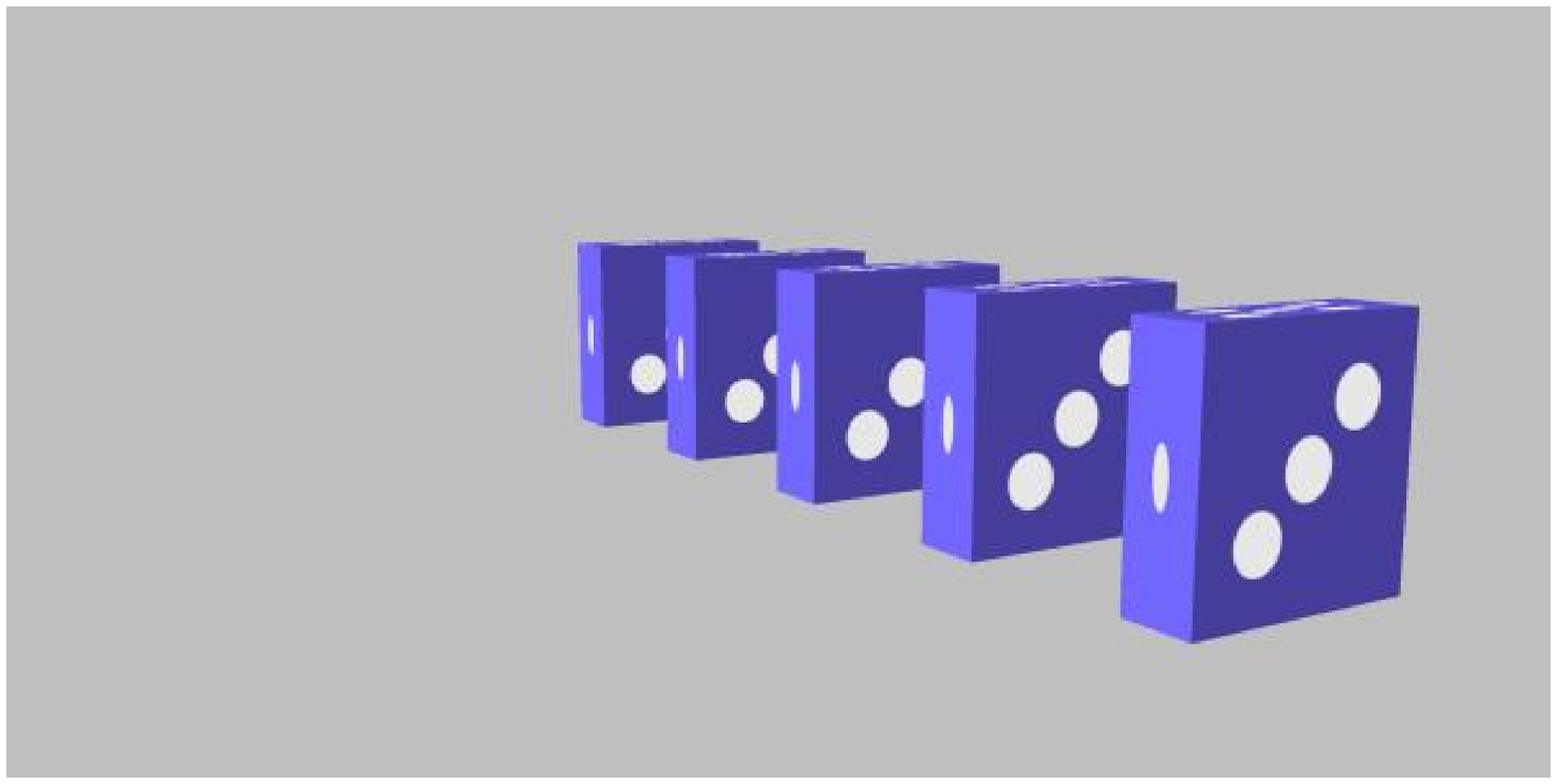,width=\cuwi}

%(c)\quad\epsfig{file=\bilder/w_1_2.ps,width=\cuwi}
(c)\quad\epsfig{file=\bilder/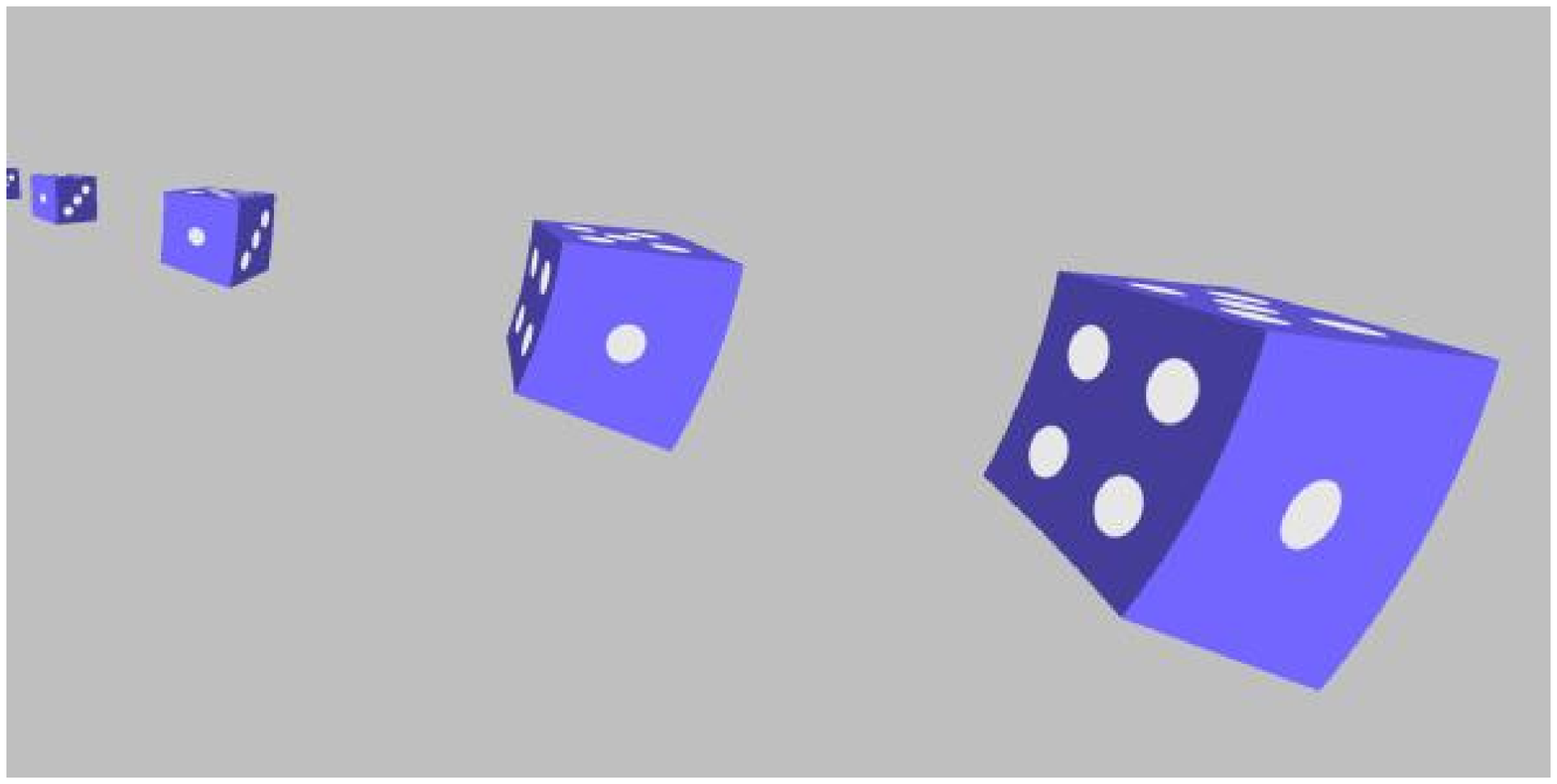,width=\cuwi}

%(d)\quad\epsfig{file=\bilder/w_2_2.ps,width=\cuwi}
(d)\quad\epsfig{file=\bilder/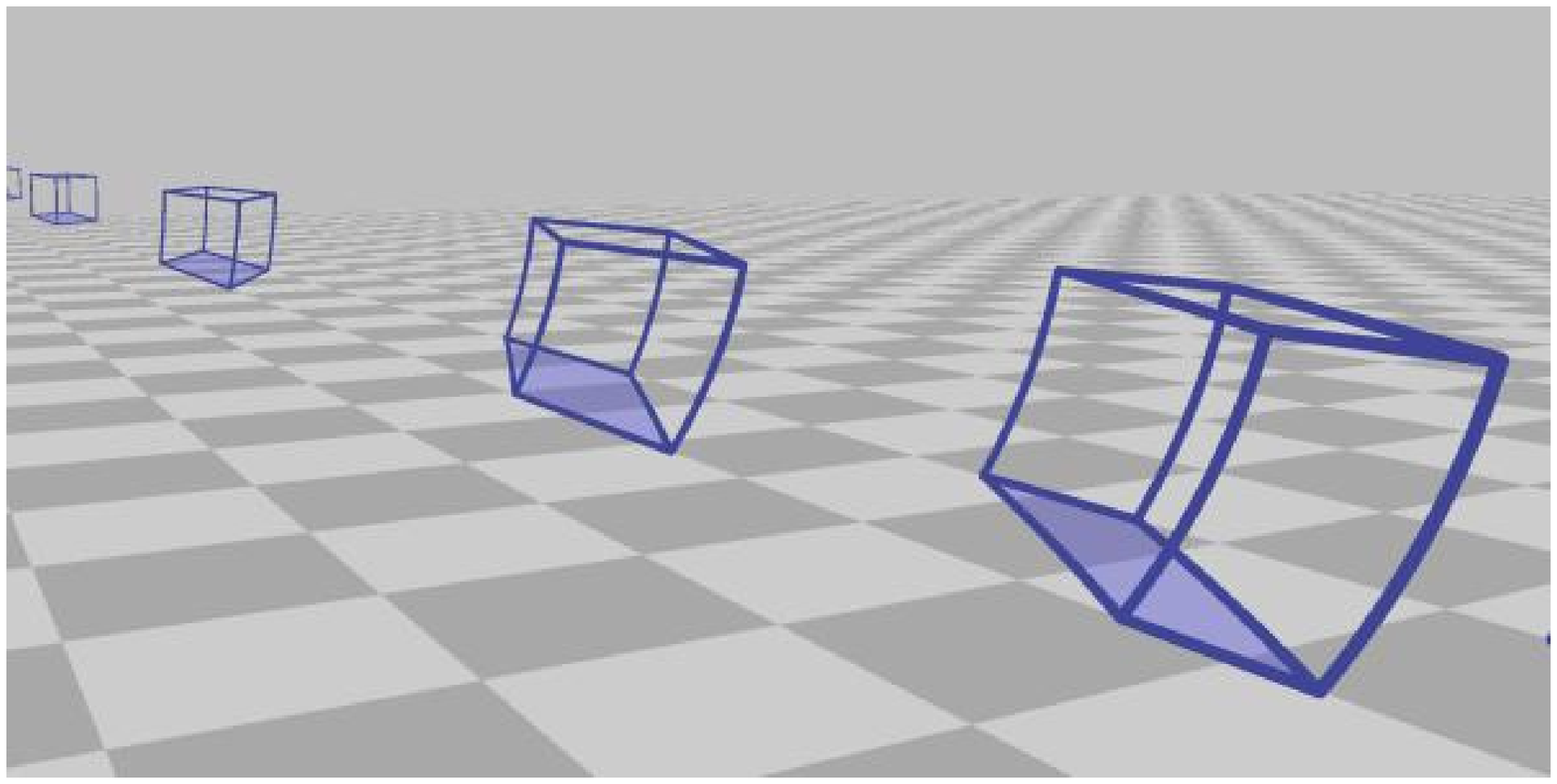,width=\cuwi}

%(e)\quad\epsfig{file=\bilder/w_4_2.ps,width=\cuwi}
(e)\quad\epsfig{file=\bilder/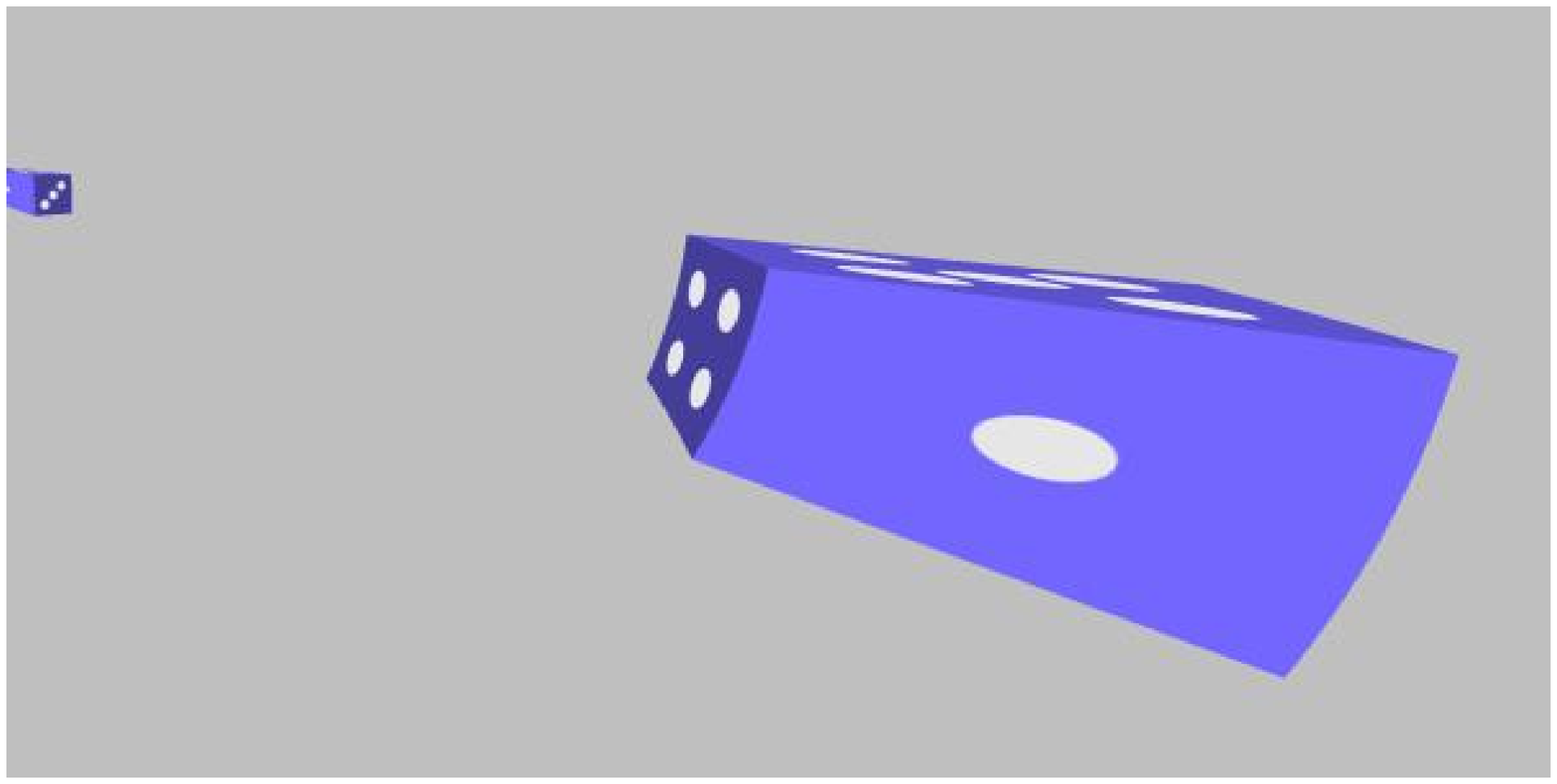,width=\cuwi}
\end{center}
\caption{\label{cube1}
A row of dice ((a), at rest) moves in a single file
at 95\% of the speed of light (motion from left to right).
The moving dice are length contracted, so that one
might (wrongly) expect them to look as shown in (b).
In the visual observation, the dice appear rotated
(c). However, when some perception in depth is
provided ((d), skeleton cubes sliding over a 
plane with their `footprints' marked) the dice
are seen as sheared rather than rotated. 
(e) The predicted `classical' appearance
of the dice (no length contraction) is shown.
(All images with a $54.4^\circ$ horizontal opening angle.)
A short movie of (c) of this figure is attached to 
the online version of the article (MPEG 1, 3 MB).
}
\end{figure}

The example shown in figure~\ref{cube1} exhibits the main
particularities of objects watched in high-speed motion.
A number of dice is set up in a row (fig.~\ref{cube1}(a), at rest).
In the simulation, these dice move in a single file 
at 95\% of the speed of light, the face with the `3' being in front
and the `4' in the rear.   
According to special relativity,
the whole arrangement suffers length contraction by the
factor $f=\sqrt{1-v^2/c^2}$, $v$~being the velocity of the dice and 
$c$~being the the speed of light; in this example $f=0.3$.
One might therefore expect to {\it see} the dice shortened
(fig.~\ref{cube1}(b)), and the early, incorrect textbook illustrations
were in fact pictures of this kind. 
The correctly computed looks of the moving dice are shown in
figure~\ref{cube1}(c). The moving dice appear primarily rotated.
Also, the vertical edges appear slightly bent, 
most notably in the dice that is closest to the observer.

\begin{figure}
\begin{center}
\epsfig{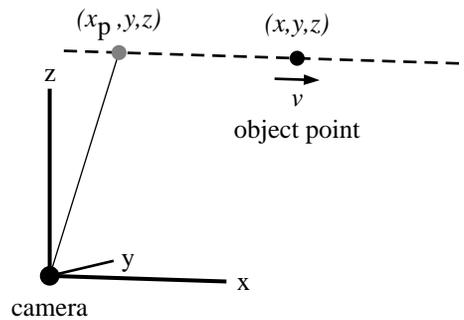}
\end{center}
\caption{\label{phantom}
A moving point source has coordinates $(x,y,z)$ at the instant
when the recording camera is released. The camera records light
that has been emitted at an earlier position $(x_ {\rm P},y,z)$
(motion is in $x$-direction).
}
\end{figure}

In this computation, the finite velocity of the light
coming from the dice has been taken into account.
As illustrated in figure~\ref{phantom}, 
when a point has coordinates $(x,y,z)$ at the time of observation,
then the light observed at that instant,
having spent a certain time to reach the observer,
must have been emitted earlier at the previous
position $(x_ {\rm P},y,z)$
(where the $x$-axis here is in the direction of motion).
A straightforward calculation shows that 
the apparent position is given by
\begin{equation}
  x_{\rm P} =  \gamma^2 x -
      \beta\gamma\sqrt{\gamma^2 x^2 + y^2 + z^2}
\label{eq_phantom}
\end{equation}
with $\beta=v/c$, 
$v$ being the velocity of the object and
$c$ being the speed of light, 
$\gamma=1/\sqrt{1-v^2/c^2}$.

The time of flight affects not only the apparent position,
but also the apparent shape of an object. 
Different points on the surface of the object 
are at different distances from the camera.
Hence, photons reaching the observer simultaneously have
not been emitted simultaneously by all points of the object.
The set of all the emission points $(x_ {\rm P},y,z)$ 
is therefore a distorted image of the object surface and what we see
is the projection of this distorted `phantom object'.

Terrell (1959) has shown that for objects subtending a small solid
angle, the  projected image
%projection of the phantom object
agrees with the projection of the suitably rotated object at rest.
This is often phrased as `a fast moving object appears rotated'.
However, this statement has been a point of some debate,
since the phantom object itself is sheared rather than rotated
and opinions differed about whether or not a snapshot would convey
the impression of shear
(Mathews and Lakshmanan 1972, Sheldon 1988, Terrell 1989).

We find that in the visualization, either impression
can be evoked. 
When the cubes are isolated objects as in figure~\ref{cube1}(c),
they clearly appear rotated. 
We can, however, make the shear visible if we
change the scene to provide a better perception
in depth. In figure~\ref{cube1}(d) the cubes slide
over a plane and, in addition, are hollow with 
the sides cut away so that only the edges remain and
the `footprints' on the plane are visible.
This image gives a visual impression of the sheared phantom cubes.
While creating these simulations, it became apparent that
it is not easy to make the cubes look sheared. 
Much guidance to the eye must be provided to dispel the
prevailing impression of rotation.

%2

In a sense, the time-of-flight effects are classical 
effects that could have been considered before
the formulation of special relativity. A comparison
with such a `classical' picture (figure~\ref{cube1}(e))
highlights the relativistic aspects: the time-of-flight
effects make the cubes appear elongated (when approaching).
Remarkably, the correct picture, including the Lorentz contraction,
is less distorted than its classical counterpart.
So, in a way special relativity makes fast looking objects
look less strange than the classical theory would.

\begin{figure}
\begin{center}
\epsfig{file=\bilder/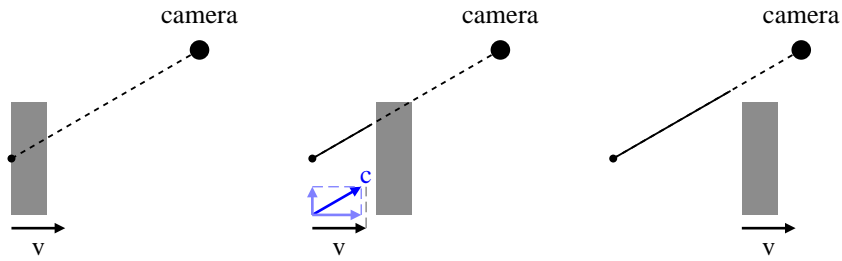}
\end{center}
\caption{\label{rear_photon}
A photon emitted from the back side of the moving cube may escape
in a forward direction, provided that the velocity component towards
the cube is less than the cube velocity.
}
\end{figure}

A particularly surprising feature of these simulations 
is the visibility of the rear side of the cubes (the `4'). 
At the first sight it seems to be impossible
for photons from the rear to reach the observer, 
since the cube is in between and should block the way. 
As is illustrated in figure~\ref{rear_photon}, the
photons escape because the cube moves out of the way fast enough.
The velocity vector of a photon directed towards the
observer can be split up into a horizontal component (towards the cube)
and a vertical component. Each component by itself is smaller than
the speed of light. The horizontal component can, therefore, be
surpassed by the cube velocity. In this case, the cube outruns
the photon and the distance between the photon and the cube increases while
the cube moves to the right and the photon towards the observer;
the photon escapes.

Visualizations of this kind can serve different purposes in
teaching. They explore consequences of the finite speed of
light that are fascinating to most students. In the context
of relativity they clearly show up the difference between visual
observations and measurements and thus 
help to point out the need for 
careful definitions of measurement procedures --
these are basic to the understanding of relativity 
but the need is not easily appreciated by many students.
The topic also lends itself to student projects:
simulations of simple objects, e.g.\ wire-frame models of
cubes, lattices or spheres, can be created with basic
programming skills. Shirer and Bartel (1967) report
on a freshman project of this kind. 
Another project that I worked on with advanced secondary school
students
is the construction of a set of real wire-frame models:
of an object, its length-contracted counterpart
and phantom objects for visual observations from different directions.
Looking at the phantom object model from the right distance then
produces the visual impression of the object in high-speed motion.
The topic is also relevant in astronomy. Motion at nearly
the speed of light is observed in some cases, most notably
quasars and gamma burst sources. Some of the visual effects that
appear in the simulations must be taken into account in order
to interpret the observations (Kraus 2005a).

%3

\subsection{Moving observer}

\begin{figure}
\begin{center}
%(a)\qquad \epsfig{file=\bilder/gassee_00.ps,width=10cm}
(a)\qquad \epsfig{file=\bilder/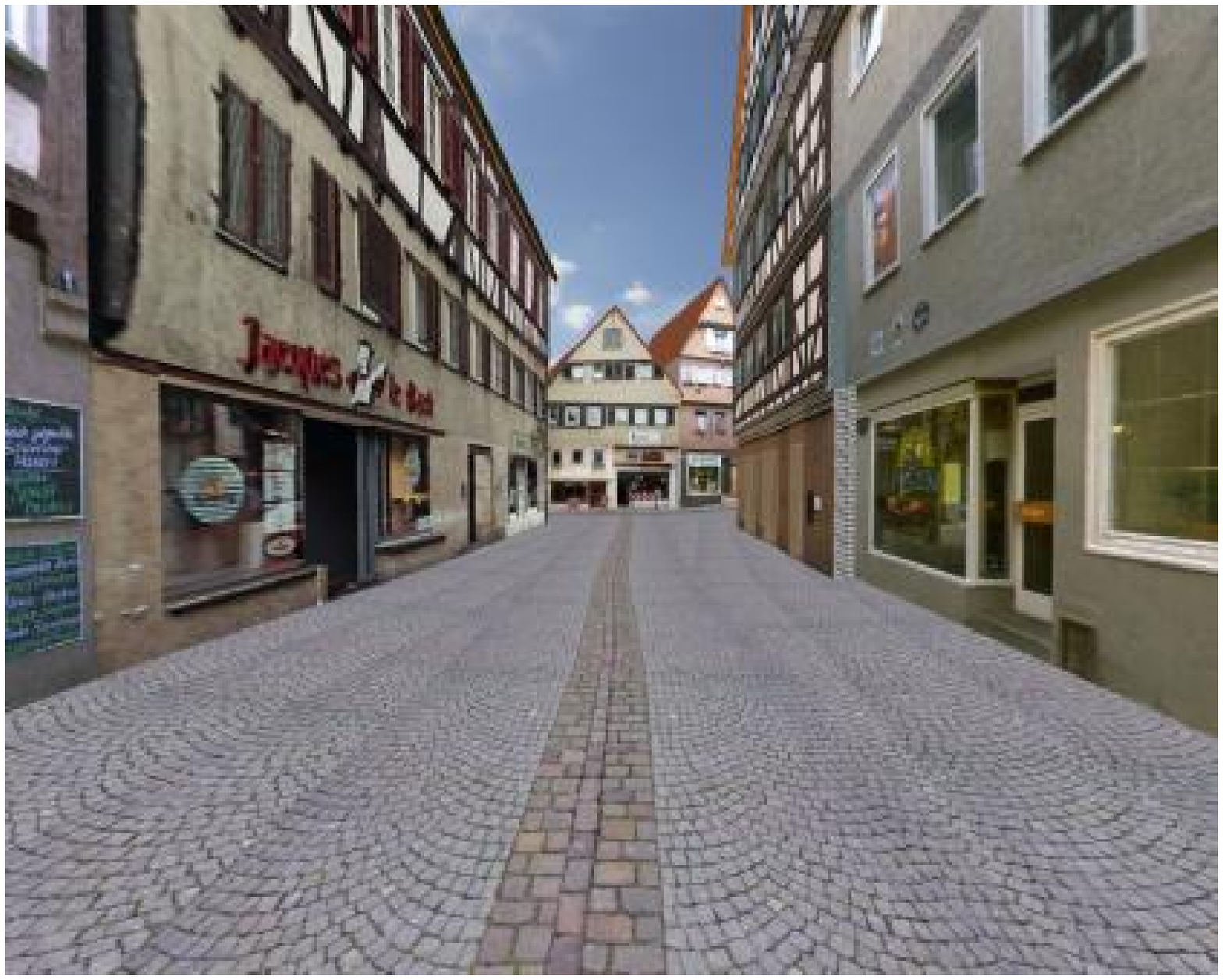,width=\tuwi}

\smallskip
%(b)\qquad \epsfig{file=\bilder/gassee_95.ps,width=10cm}
(b)\qquad \epsfig{file=\bilder/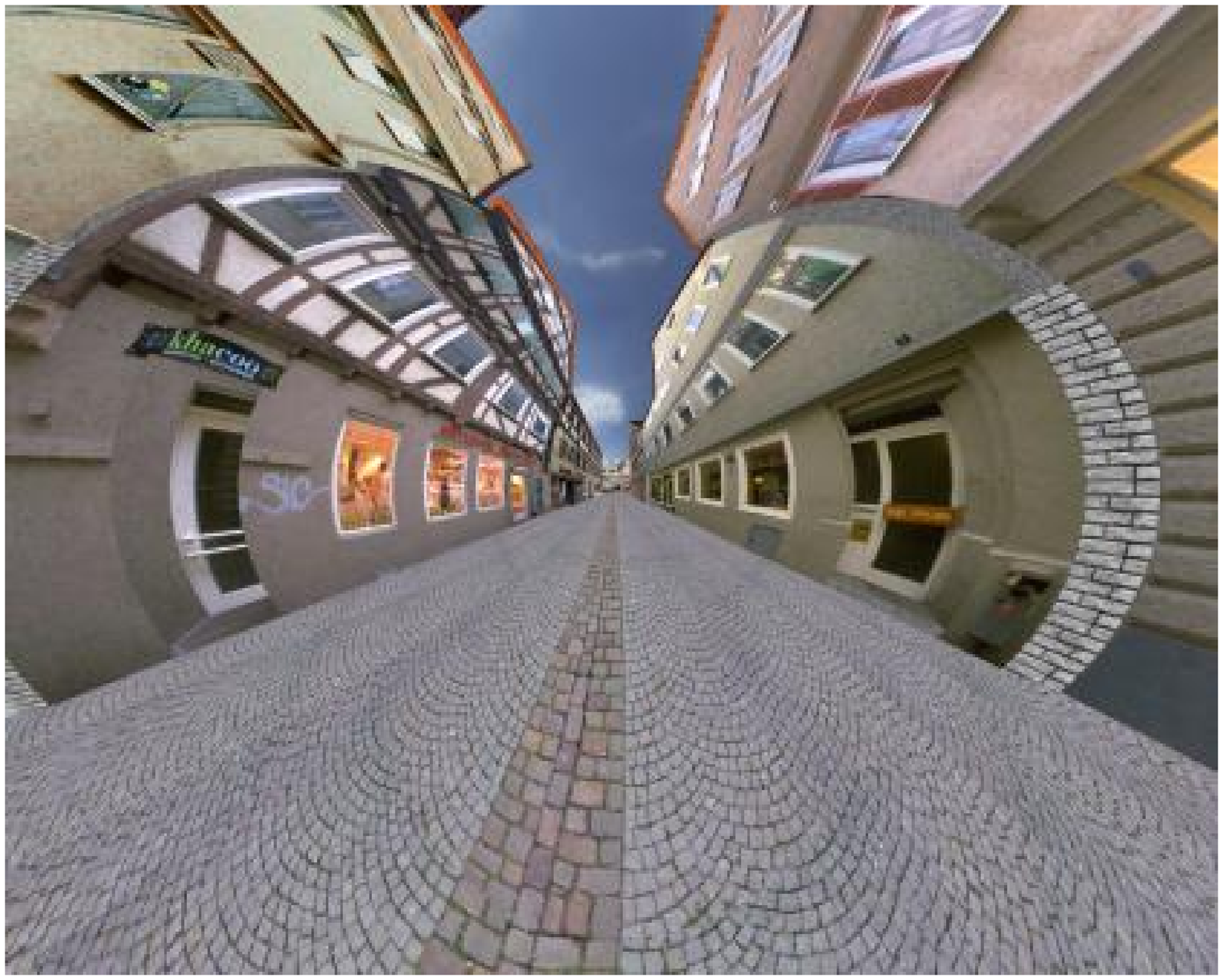,width=\tuwi}
\end{center}
\caption{\label{tuebingen}
Looking down the Kornhausstra{\ss}e towards the Hirschgasse in T\"ubingen.
(a) Snapshot while at rest.
(b) Snapshot while moving down the road at 95\% of the speed of light,
   looking forward.
Both snapshots are taken {\em at the same place}.
The opening angle of the camera is $90^\circ \times 112.5^\circ$,
i.e.\ these are wide-angle images. In order to obtain
the visual impression that one would have with the naked eye,
one would have to scale up the picture to DIN A1 format
and then look at it from a distance of 30~cm.
Computer simulation by Marc Borchers.
For two movies of this simulation, see the online version of
Kraus and Borchers (2005).
}
\end{figure}

The visual perception of a moving observer is best
demonstrated in comparison with an observer at rest {\em at the
same place}.
Figure~\ref{tuebingen} shows two snapshots taken from a simulation
using as setting a detailed three-dimensional model of the old city 
centre of T\"ubingen\footnote{
We thank Professor H B\"ulthoff for his kind permission to use this model
that has been constructed at the Max Planck Institute for Biological
Cybernetics, T\"ubingen.%
}.
In the simulation, the speed of light in `virtual T\"ubingen' 
is reduced so that we experience relativistic effects while riding a bike
through the city (Borchers 2005, Kraus and Borchers 2005).

%4

Figure~\ref{tuebingen}(a) is a snapshot taken while standing
in the Kornhausstra{\ss}e and looking down the street towards
the Hirschgasse. Note the bakery on the left hand side 
and the ice cream parlour at the end of the road. 

The second snapshot, figure~\ref{tuebingen}(b), is taken 
while driving down the road at 95\% of the speed of light. 
The camera is released at the instant when we pass the spot
where the first snapshot was taken. We now see the bakery
far in front of us, the ice cream parlour at the end of the
road is hardly visible and close to us the edges of the houses
appear bent. Next to the bakery, two more houses appear in
the picture which are actually standing to our side and
behind us.

Aberration is the reason for the differences between the two snapshots.
A light ray that reaches a stationary observer at an angle $\theta$
to the direction of motion (where $\theta=0$ is a ray arriving head on)
is perceived by the moving observer at an angle $\theta'$ with
\begin{equation}
\cos\theta'=\frac{\cos\theta+\beta}{1+\beta\cos\theta},
\label{eq_aberration}
\end{equation}
where $\beta=v/c$, $v$ being the velocity of the oberserver and
$c$ being the speed of light.

\begin{figure}
\begin{center}
\epsfig{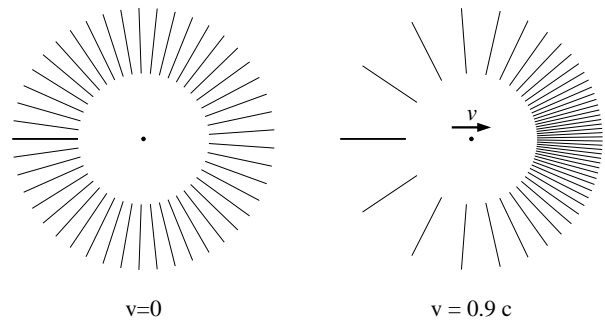}
\end{center}
\caption{\label{fig_aberration}
Aberration.
Left: a number of light rays reaching a stationary observer evenly
from all sides. Right: a moving observer at the same place attributes
different direction to the same light rays.}
\end{figure}

Figure~\ref{fig_aberration} illustrates this effect.
A number of light rays is shown that reach the stationary observer
evenly from all sides. The directions that the moving observer 
attributes to these same light rays are shifted towards the front.
She therefore perceives objects in front of her within a smaller
solid angle which makes them appear to be farther away.
Also, light rays which reach the stationary observer obliquely
from behind may to the moving observer come obliquely from her front;
for this reason, a high-speed camera can image objects 
located behind its back.

\begin{figure}
\begin{center}
\epsfig{file=\bilder/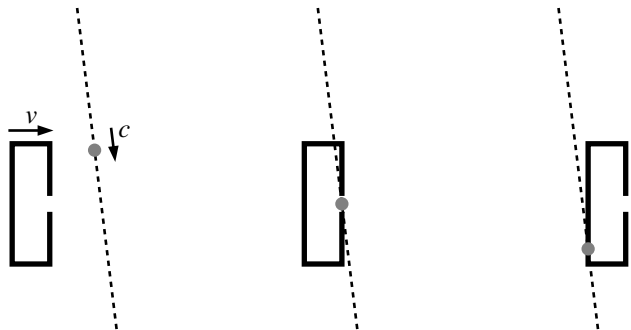}
\end{center}
\caption{\label{camera}
A moving camera can image objects located behind its back.
The pinhole camera moves from left to right at 95\% of the speed of light.
A photon approaching from above and behind the camera is 
caught by the pinhole and then swept up by the image plane.
}
\end{figure}

One may ask how precisely light that
is coming obliquely from behind is supposed to enter the moving
camera that is looking forward all the time.
The answer is illustrated 
in figure~\ref{camera} with a pinhole camera:
a photon approaching obliquely from behind moves in front of the
camera, is caught by the pinhole and is then swept up by the image plane.

Apart from snapshots and movies, the T\"ubingen simulation 
also allows interactive manipulation.
The user determines speed and direction, while the apparent
view of the city is displayed in real time. 
The interactive control can be exercised via a joystick. 
There is also an `exhibition version' simulating a bike ride, 
where the user sits on an exercise bike. Turns of the handle bar and
pedalling determine direction and speed respectively. 
The apparent view of the
street scene is projected onto a large screen in front of the bike.

There are features of the simulation that 
are best experienced in an interactive form,
for example the effect of acceleration. Whenever one accelerates, the
houses in front seem to move into the distance; this is reversed
upon slowing down. The reason is that with increasing velocity,
the aberration becomes stronger and objects seen in front 
take up a smaller solid angle. 
Because of looking smaller, they seem to be farther
away.
The use of a similar interactive simulation in an undergraduate
relativity class has been described by Savage et al. (2007).

%5

%6

\begin{figure}
\begin{center}
%(a)\quad\epsfig{file=\bilder/w_0_3.ps,width=\cuwitwo}
(a)\quad\epsfig{file=\bilder/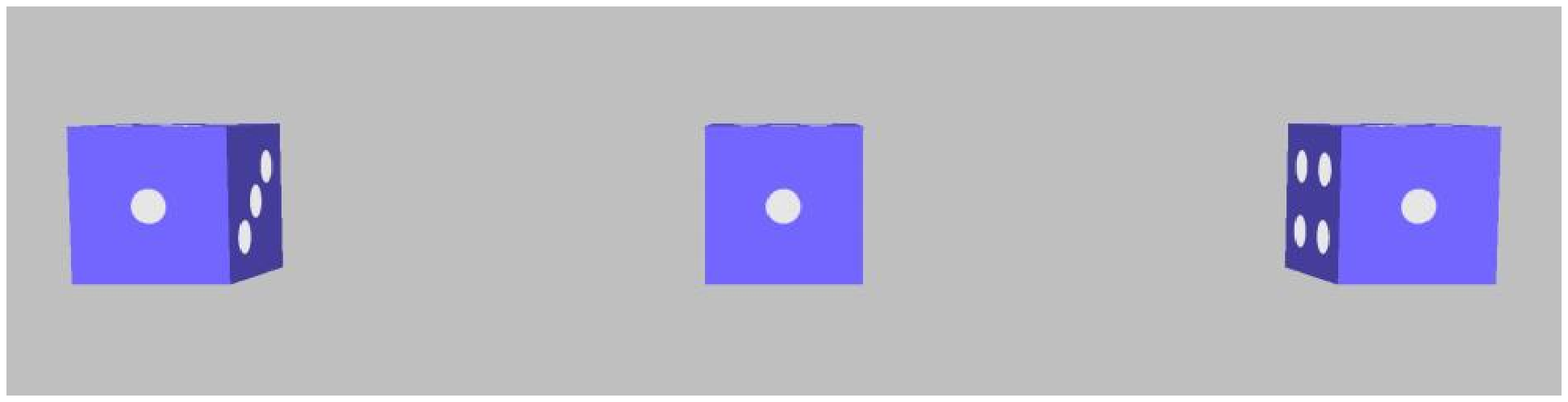,width=\cuwitwo}

%(b)\quad\epsfig{file=\bilder/w_1_3.ps,width=\cuwitwo}
(b)\quad\epsfig{file=\bilder/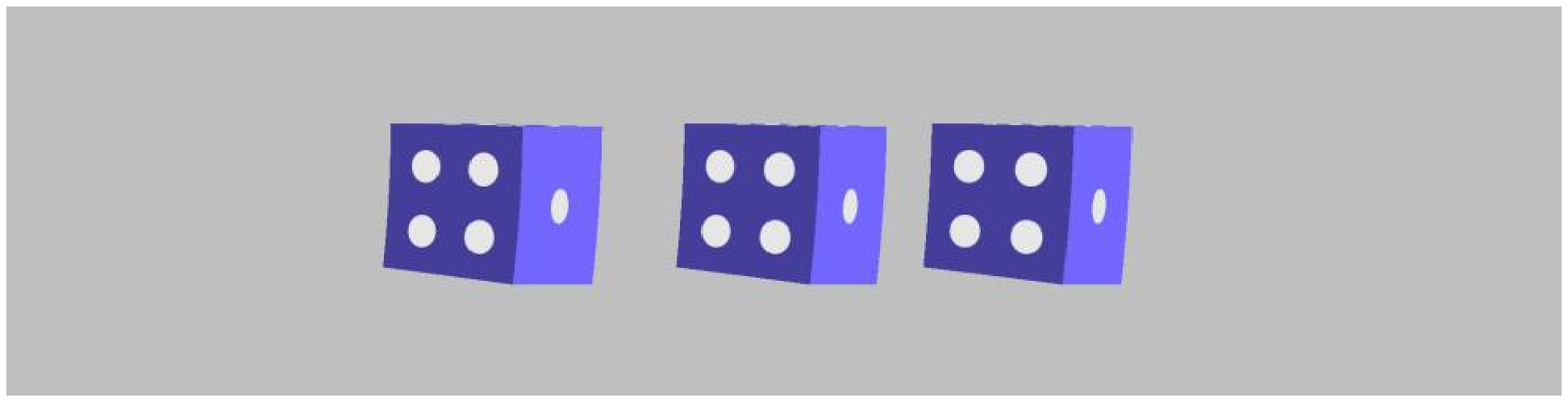,width=\cuwitwo}

%(c)\quad\epsfig{file=\bilder/w_4_3.ps,width=\cuwitwo}
(c)\quad\epsfig{file=\bilder/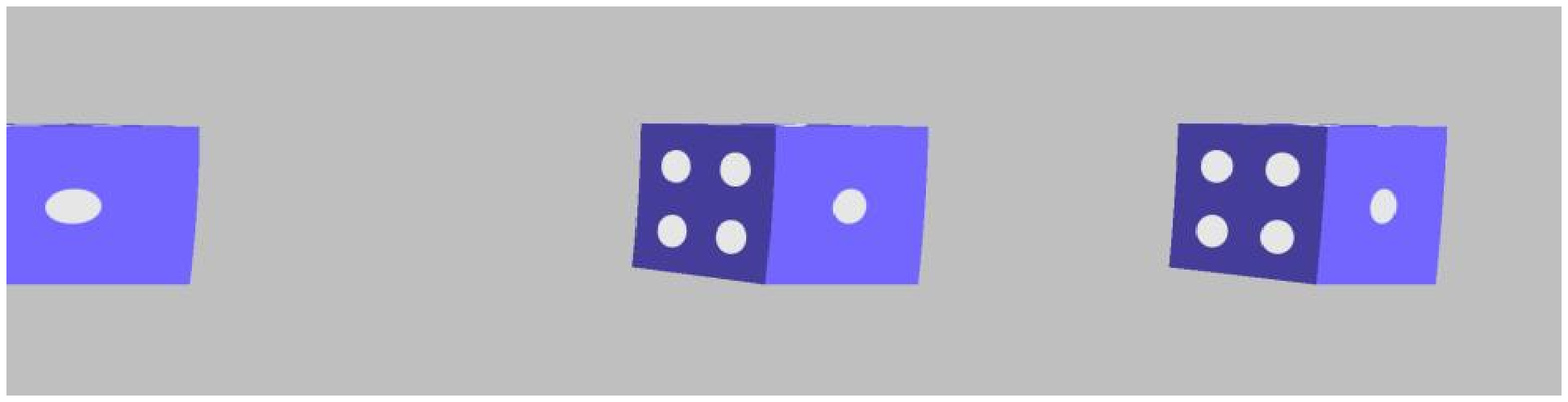,width=\cuwitwo}

%(d)\quad\epsfig{file=\bilder/w_5_3.ps,width=\cuwitwo}
(d)\quad\epsfig{file=\bilder/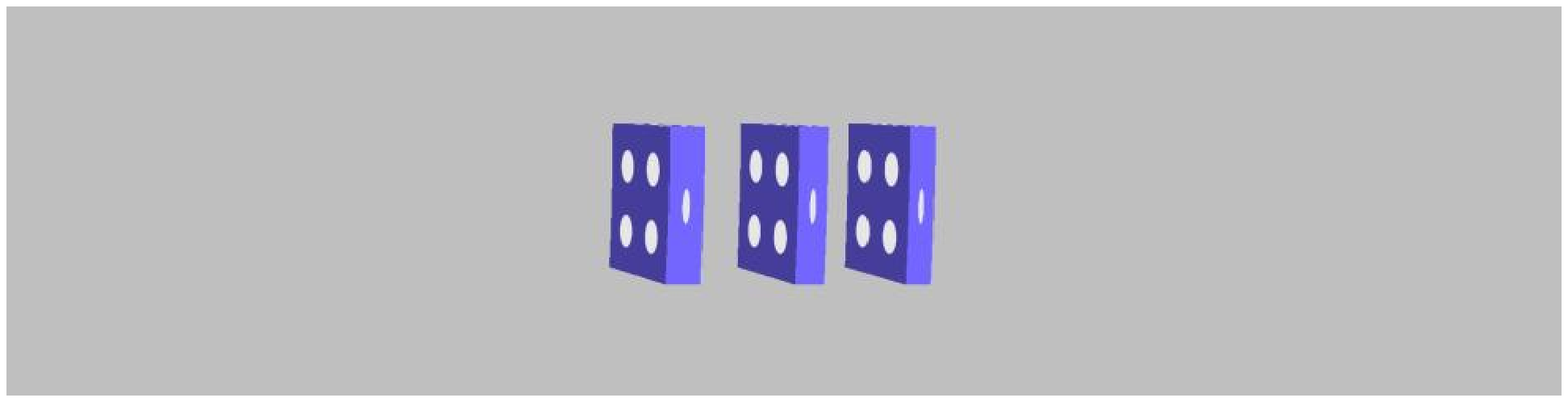,width=\cuwitwo}
\end{center}
\caption{\label{cube2}
Three dice ((a), at rest) are watched at 
a relative velocity of 90\% of the speed of light.
The direction of motion is perpendicular to the
line of sight.
(b) The dice moving from left to right or the observer
from right to left, the visual appearance is the same.
(c) `Classical' computation of the dice moving from left to right
(no length contraction),
(d) ``Classical'' computation for an observer moving
from right to left (non-relativistic aberration formula).
(All images with a $54.4^\circ$ horizontal opening angle.)
}
\end{figure}

The visualizations of high-speed motion can be used to illustrate
the principle of relativity.
A snapshot taken at high relative velocity between the object
and the observer may be computed in one of two ways: either
the object is considered to be moving and its appearance is
the projection of the respective phantom object or the observer
is considered to be moving and his visual perception is determined
by applying the aberration formula. Both methods give the same
result as one can deduce from equations~(\ref{eq_phantom}) 
and~(\ref{eq_aberration}).
It is instructive to compare this to the non-relativistic case.
Figure~\ref{cube2} shows again a row of dice ((a), at rest),
here seen with the line of sight perpendicular to the direction of motion.
In the correctly computed image, the dice appear rotated
(figure~\ref{cube2}(b)). These may either be moving dice watched
by a stationary observer or vice versa. 
The `classical' moving dice (figure~\ref{cube2}(c)),
computed without the Lorentz contraction, appear both
rotated and elongated.
In contrast, the `classical' moving observer whose snapshot
is computed using the non-relativistic aberration formula
sees the dice rotated and contracted (figure~\ref{cube2}(d))!

The simulation of a high-speed observer's perception is
a programming project that is not too difficult for
a simple scene consisting e.g.\ of geometric objects
(for a sample programme see Kraus and Zahn (2003)).

These visualizations may also be of interest in teaching astronomy.
Aberration is an important phenomenon in astronomy, and the 
visual observations of a fast moving observer provide
impressive illustrations.

%7

\section{Visualizations of the general theory of relativity}

The general theory of relativity is a theory of gravitation.
First-person visualizations in this context may explore
visual observations in high-gravity surroundings.
Typical examples are observations near neutron stars or black holes
as in the early work by Luminet (1979) and Nemiroff (1993).
In the vicinity of compact objects,
gravitational light deflection is strong and generates surprising
visual effects.

%8
\begin{figure}
\begin{center}
%(a)\quad\epsfig{file=\bilder/panorama_ax4.ps,width=\bhwi}
(a)\quad\epsfig{file=\bilder/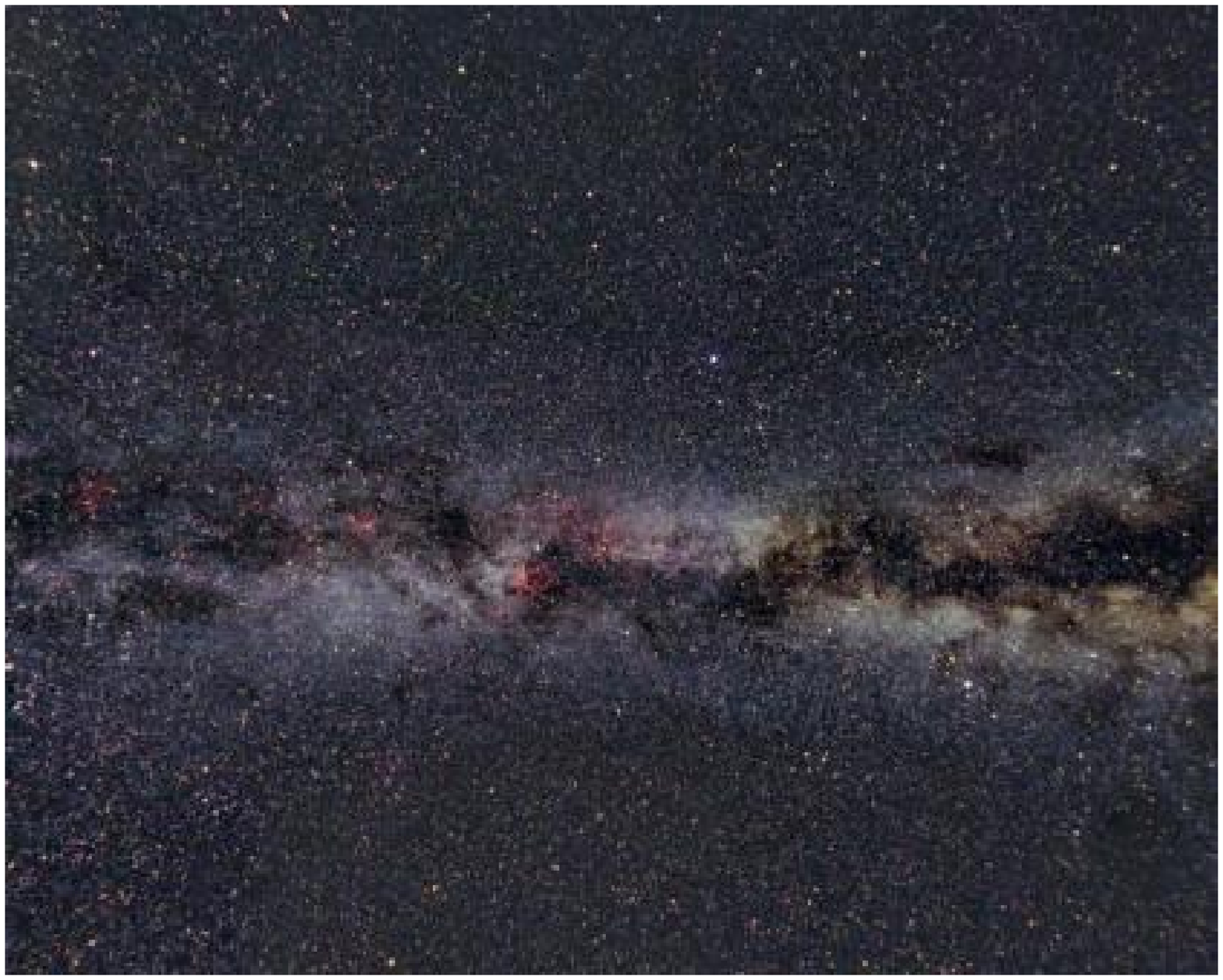,width=\bhwi}

%(b)\quad\epsfig{file=\bilder/s0_4ax4.ps,width=\bhwi}
(b)\quad\epsfig{file=\bilder/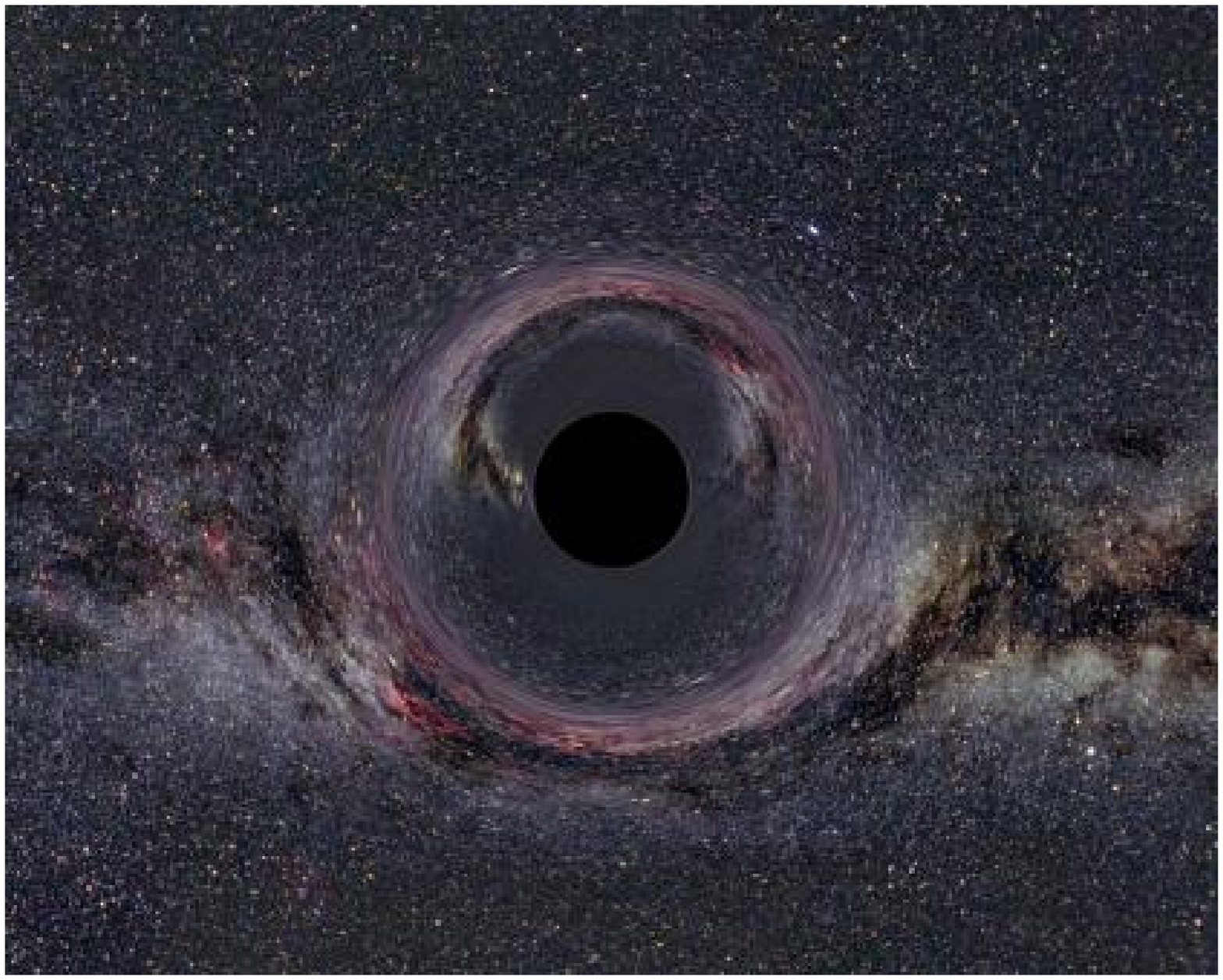,width=\bhwi}

\end{center}
\caption{\label{blackhole}
A black hole of ten solar masses
in front of the Milky Way (a), seen from a 
distance of 600~km (b), 45~km (c) and 13~km (d)
above the event horizon.
In (b) we look towards the black hole, in (c) 
to the side, in (d) away from it.
(All images with a $90^\circ$ horizontal opening angle.)
}
\end{figure}

\begin{figure}
\begin{center}
%(c)\quad\epsfig{file=\bilder/s0_03bx4.ps,width=\bhwi}
(c)\quad\epsfig{file=\bilder/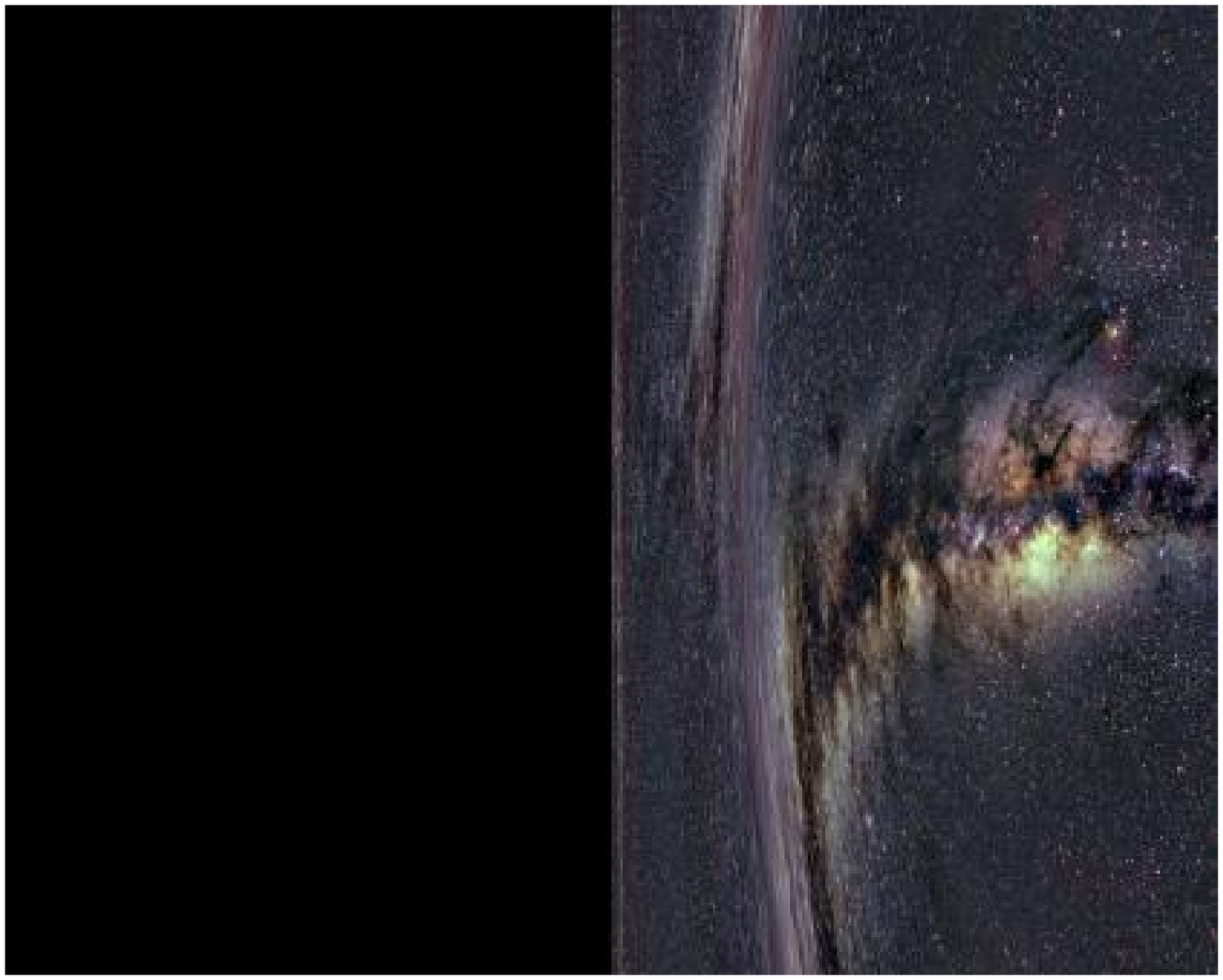,width=\bhwi}

%(d)\quad\epsfig{file=\bilder/s0_021cx4.ps,width=\bhwi}
(d)\quad\epsfig{file=\bilder/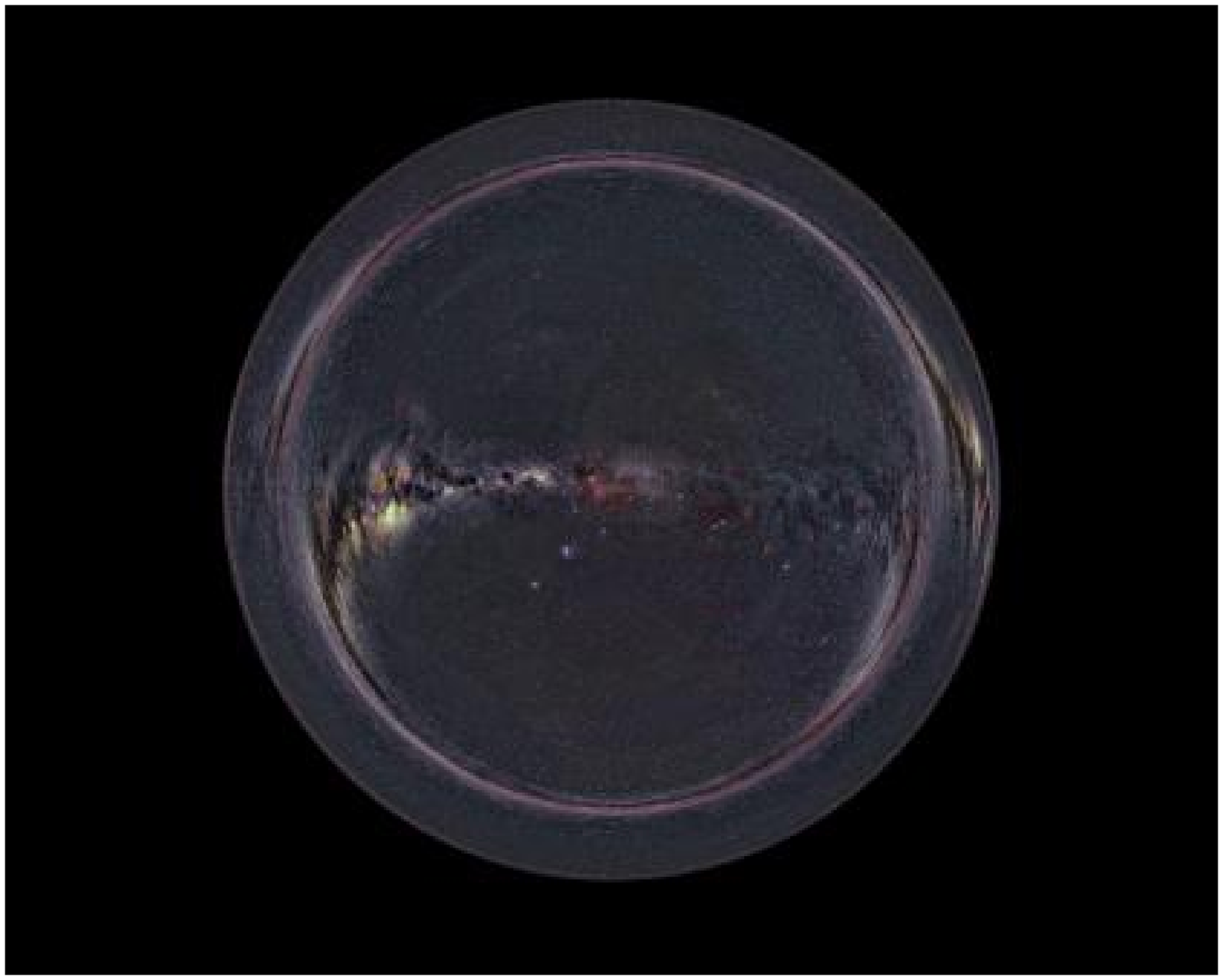,width=\bhwi}

\medskip
{\small Figure 8} {\small (Continued.)}\hspace*{5.5cm}

\end{center}
\end{figure}

To illustrate this, figure~\ref{blackhole} shows some snapshots
from a virtual journey towards a black hole (Kraus 2005b).
The numbers given apply to a black hole of ten solar masses
(Schwarzschild radius $\rs=30\,\rm km$). 

In the visualization, we place the black hole in front of
the Milky Way%
\footnote{
This background image is the All-Sky Milky Way Panorama by
Axel Mellinger (Mellinger 2000, 2005).
} (figure~\ref{blackhole}(a)).
Each snapshot is taken while the observer keeps his distance to 
the black hole constant, i.e.\ while he is hovering near the
black hole. 
From a distance of 600~km (radial coordinate $r=20\,\rs$),
the Milky Way in the background appears distorted with arc-like
structures and double images (figure~\ref{blackhole}(b)) close to
the central black disc from which no light is received. 
Farther in, the black disc gets larger and larger 
until it completely fills the front view. 
The camera is then turned to the side.
45 km above the event horizon ($r=1.5\,\rs$) it
records the image of figure~\ref{blackhole}(c), in which the black disc
covers exactly half the sky. 
Moving further in, the black disc grows still larger and
finally fills the side view as well. The camera is then turned to
the back and at 13~km above the horizon ($r=1.05\,\rs$) records
figure~\ref{blackhole}(d): the observer seems to be engulfed by
the black hole, the night sky is reduced to a small
round patch, diminishing ever more as the observer approaches
the horizon of the black hole.

In these simulations, the photon paths $x^\mu(\lambda)$
from the background image
to the observer are computed as solutions of the geodesic equation
\begin{equation}
\frac{d^2x^\mu}{d\lambda^2}+
\sum_{\nu,\kappa=0}^3
\Gamma_{\nu\kappa}^\mu \, \frac{dx^\nu}{d\lambda}\frac{dx^\kappa}{d\lambda} =0 
\qquad \mbox{subject to} \qquad
\sum_{\mu,\nu=0}^3 
g_{\mu\nu}\frac{dx^\mu}{d\lambda}\frac{dx^\nu}{d\lambda} = 0
\label{geodesic}
\end{equation}
with the spacetime metric $g_{\mu\nu}$
and the Christoffel symbols 
$\Gamma_{\nu\kappa}^\mu = 0.5 \sum_{\sigma=0}^3 g^{\mu\sigma} 
  (\partial g_{\nu\sigma}/\partial x^\kappa + 
   \partial g_{\kappa\sigma}/\partial x^\nu  -
   \partial g_{\nu\kappa}/\partial x^\sigma)$.
This is a set of four coupled ordinary differential equations
which is solved numerically.

%9
\begin{figure}
\begin{center}
\epsfig{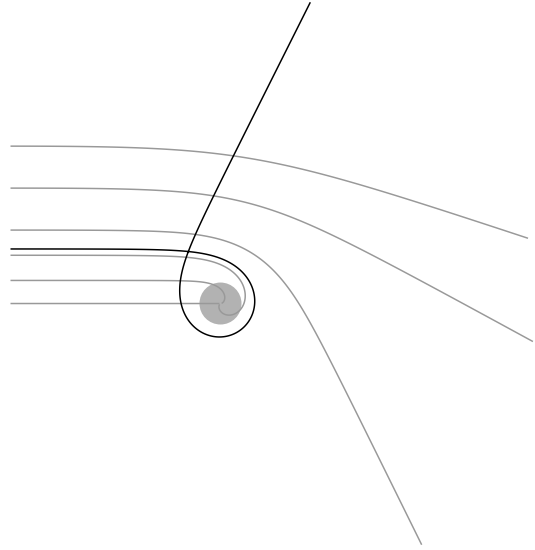}
\end{center}
\caption{\label{orbits}
Photon orbits near a black hole. The region $r<\rs$ ($\rs$ the
Schwarzschild radius)
is shaded in grey. Close to the photon radius $1.5\,\rs$, 
the deflection may become very large.
}
\end{figure}

Close to a black hole, gravitational light deflection is
a dramatic effect as is illustrated in figure~\ref{orbits};
near the photon radius at radial coordinate $r=1.5\,\rs$, 
photons are so strongly deflected that they may circle the black hole,
even several times, before either escaping or crossing the event horizon.
The photon radius itself is an instable circular photon orbit 
that forms the borderline between photon orbits that
extend to infinity and those that end up in the black hole.

This explains the peculiar observation that the black hole appears
to cover half the sky (figure~\ref{blackhole}(c)) when
the observer is still some way outside the horizon.
The image is a snapshot taken when the observer is located 
on the photon radius.
In this position, when looking exactly to the side ($90^\circ$ to the
radial direction) the observer receives photons moving on the circular orbit,
i.e.\ coming from his own back of the head. 
Only photons that arrive from outside of the circular orbit
come from the exterior of the black hole. 
The exterior region, therefore, covers exactly half the
observer's sky and the black disc the other half.

For teaching purposes, an interactive black hole simulation 
offers the possibility of experimenting with light deflection
(Zahn and Kraus 2006). In this simulation,
a black hole is placed in between the observer and a background image,
with a large distance to both, and the user can move the black hole
across the image interactively. This may be used to demonstrate
several astrophysically important effects:
changes in the apparent position (as observed in stars near the sun 
during a solar eclipse, one of the classical tests
of general relativity), arc-like deformations (as observed in 
gravitational lensing of galaxies)
and the formation of an Einstein ring.
The simulation also clearly shows the
large increase in brightness when an Einstein ring is formed,
an effect which is exploited to search for dark matter 
in the Milky Way in the shape of compact halo objects
(Paczy\'nski 1996).

In courses on general relativity, 
photon orbits in a Schwarzschild spacetime are a standard topic 
and visualizations such as those described above 
may be used as complements to the usual analytical treatment.

In astronomy, gravitational light deflection has increasingly
become an important topic. While at the beginning of the twentieth century
it seemed to be an exotic phenomenon,
observed in order to test a new theory of gravitation,
it is now a tool used routinely in the study of cosmic objects
ranging from extrasolar planets to the structure of galaxies.
Visualizations can illustrate many of these astronomical applications
(Kraus 2006).
 
All the visualizations shown in this contribution focus on geometric
effects. In order to make the geometry clearly visible, Doppler
and searchlight effects have not been taken into account. 
However, these effects are dramatic in both high-speed motion
and strong gravity environments (Kraus 2005a). In order to include them,
the spectrum of the objects must be known, not only in the visible
range but also beyond. To compute the observed colour from the
suitably transformed spectra is an unusual and interesting application
of colorimetry (Kraus 2000).

\section{Teaching use}

Visualizations of relativistic effects are a complement to
the standard approaches to the special and general theory of
relativity. The pictures and movies serve a number of purposes:
as an introduction, they are sure to arouse interest. We have
found that objects and scenes from everyday life work best
in this respect. 
Secondly, by showing extreme situations, 
the images illustrate effects in a dramatic
and highly visible way, aiding long-term memory.
Thirdly, and maybe most importantly, they act as a counterbalance
to wrong or mistaken pictures and concepts. Examples are widely used
pictures of black holes (Berry 2001, 2003)
or common misinterpretations of terms like `observing' which
many students readily equate with `observing visually'. 
Images that are both impressive and correct are the best
means to replace misconceptions of this kind with more correct notions.

In undergraduate education, we consider these visualizations
to be especially useful when teaching relativistic physics to 
students who will continue as physics teachers, because our 
experience shows that this material can very profitably be utilized in
secondary schools. The travelling exhibition `Einsteinmobil'
(see http://www.einsteinmobil.de for details)
has been bringing interactive simulations and movies into
secondary schools since the beginning of 2006. The
exhibits stay for a typical duration of 2~weeks and are
used at the discretion of the school. To evaluate the project,
teachers of the participating schools are asked to complete
questionnaires. A general finding is that most or all
students show much interest, independent of age, of gender and
of their general interest in physics. 

The visualizations of our group are being used 
in a wide range of teaching situations
from undergraduate university classes and physics teachers seminars
to secondary school projects, planetarium shows and museum exhibitions.
Vollmer (2006) even describes a project in primary school.

In order to make the visualizations available to both instructors and students,
we maintain the web site http://www.spacetimetravel.org
(with its German counterpart\\
 http://www.tempolimit-lichtgeschwindigkeit.de)
on which we present a collection of multimedia contributions, 
including many short simulation movies. 
The contributions range from
entertaining introductions (e.g.\ relativistic soccer)
to more theoretical papers on an advanced undergraduate level. 
For use by instructors, 
the simulation movies are provided in a quality suitable for 
projection in class.
This site will be updated with future visualization projects.

\small
\section*{References}
%\begin{harvard}
\begin{description}
\renewcommand{\itemsep}{0.cm}
\item[] Berry 2001 
        http://sci.esa.int/science-e/www/object/index.cfm?fobjectid=28795
\item[] Berry 2003 
        http://sci.esa.int/science-e/www/object/index.cfm?fobjectid=29886
\item[] Borchers M 2005 Interaktive und stereoskopische Visualisierung in
        der speziellen Relativit\"atstheorie {\it Dissertation}
        Universit\"{a}t T\"{u}bingen. English summary and German full text:\\
        http://w210.ub.uni-tuebingen.de/dbt/volltexte/2005/1891/
\item[] Fuchs W 1965 {\it Knaurs Buch der Modernen Physik} 
        (M\"unchen: Droemer Knaur) p 228
\item[] Gamow G 1940 {\it Mr Tompkins in Wonderland}
        (Cambridge: Cambridge University Press)
        (reprint 1993 {\it Mr Tompkins in Paperback}) 
\item[] Gamow G 1961 {\it Proc. Natl Acad. Sci.}
        {\bf 47} 728
\item[] Kraus U 2000 {\it Am. J. Phys.} {\bf 68} 56
\item[] Kraus U 2005a {\it Sterne Weltraum} {\bf 8/2005} 40
        Online version in English: {\it Motion near the cosmic speed limit}
        on http://www.spacetimetravel.org
\item[] Kraus U 2005b {\it Sterne Weltraum} {\bf 11/2005} 46
        Online version in English (abbreviated):
        {\it Step by step into a black hole}
        on http://www.spacetimetravel.org
\item[] Kraus U 2006 {\it Sterne Weltraum} {\bf 10/2006} 38
        Online version: {\it R\"ontgenpulsare} on\\
        http://www.tempolimit-lichtgeschwindigkeit.de
\item[] Kraus U and Borchers M 2005 {\it Phys.\ in unserer Zeit} 
        {\bf 2/2005} 64 Online version in English: 
        {\it Through the city at nearly
        the speed of light} on http://www.spacetimetravel.org,
        online version in German:
        {\it Fast lichtschnell durch die Stadt} on
        http://www.tempolimit-lichtgeschwindigkeit.de
\item[] Kraus U and Zahn C 2003 {\it Relativistic Flight Through a Lattice}
        on http://www.spacetimetravel.org
\item[] Lampa A 1924 {\it Z. Phys.} {\bf 27} 138
\item[] Luminet J-P 1979 {\it Astron. Astrophys.} {\bf 75} 228
\item[] Mathews P and Lakshmanan M 1972 {\it \NC} {\bf 12} 168
\item[] Mellinger A 2000 {\it Sterne Weltraum} {\bf 2-3/2000} 174
\item[] Mellinger A 2005 http://home.arcor-online.de/axel.mellinger
\item[] Nemiroff R 1993 {\it Am. J. Phys.} {\bf 61} 619
\item[] Paczy\'nski B 1996 {\it Ann. Rev. Astron. Astrophys.} {\bf 34} 419
\item[] Penrose R 1959 {\it Proc. Camb. Phil. Soc.} {\bf 55} 137
\item[] Savage C, Searle A and McCalman L 2007 {\it Preprint} physics/0701200v1
\item[] Sexl R 1980 {\it Naturwissenschaften} {\bf 67} 209
\item[] Sheldon E 1988 {\it Am. J. Phys.} {\bf 56} 199
\item[] Shirer D and Bartel T 1967 {\it Am. J. Phys.} {\bf 35} 434
\item[] Terrell J 1959 {\it \PR} {\bf 116} 1041
\item[] Terrell J 1989 {\it Am. J. Phys.} {\bf 57} 9
\item[] Vollmer M 2006 {\it Proc. DPG-Tagung Physikdidaktik Kassel 2006}
        (CD-ROM)
\item[] Zahn C and Kraus U 2006 {\it Interactive black hole simulation}
        on http://www.spacetimetravel.org
\end{description}
%\end{harvard}

\end{document}